\documentclass[twocolumn,showpacs,superscriptaddress]{revtex4}
\usepackage{graphicx}
\usepackage{amsmath}
\usepackage{bm}% bold math

\begin{document}

\title{Coupled uncertainty provided by a multifractal random walker}

\author{Z. Koohi Lai $^{1}$, S. Vasheghani Farahani$^2$, S.M.S. Movahed $^{3,4}$, G.R. Jafari $^{3}$ \thanks{Email: g\_jafari@sbu.ac.ir}$^\dag$\\
{\small $^1$ Department of Physics, Firoozkooh Branch, Islamic Azad university, Firoozkooh, Iran } \\
{\small $^2$ Department of Physics, Tafresh University, P.O. Box 39518-79611, Tafresh, Iran} \\
{\small $^3$ Department of Physics, Shahid Beheshti University,
G.C., Evin, Tehran 19839, Iran} \\
{\small $^4$ The Abdus Salam International Centre for Theoretical Physics, Strada Costiera, 11,  Trieste 34151, Italy} \\
{\small Corresponding author: \ \ g\_jafari@sbu.ac.ir} \\
 }
\date{\today}

\begin{abstract}
The aim here is to study the concept of pairing multifractality
between time series possessing non-Gaussian distributions. %A rare
%event in a Non-Gaussian distribution is more likely to occur
%compared with its corresponding in a normal distribution.
The increasing number of rare events creates "criticality".
We show how the pairing between two series is affected by rare events, which we
call "coupled criticality". A method is proposed for studying the coupled criticality born
out of the interaction between two series, using the bivariate multifractal random walk
(BiMRW). This method allows studying dependence of the
coupled criticality on the criticality of each individual system.
%MRW systems which would eventually create an anomalous coupling. A
%measure for the joint-Gaussianity of the probability density
%function represented by $\Lambda_{\ell}$ for two underlying systems at
%scale $l$ is presented. %This measure is obtained based on the
%reciprocal multiplicative cascade notion in the multifractal
%formalism. A large value of $\Lambda_{\ell}$ refers to a robust coupled
%multifractality resulting in a coupled criticality in the series.
%We show how the coupled criticality is highly dependant on the
%individual criticality of the systems. but interestingly, it is
%shown that in a special case where the two series individually
%posses a Gaussian distribution, long-ranged cross correlations
%become present. This causes the coupling of underlying series to be
%in its strongest state. This state is named here as the 'state of
%resonance' in which the coupled criticality is extreme.
This approach is applied to data sets of gold \& oil markets, and inflation \& unemployment.
\end{abstract}

\pacs{02.50.-r,05.40.Fb,89.65.Gh,89.75.Da}

\maketitle

%\keywords{Coupling, multifractal random walk, criticality }

\section{introduction}
The concept of coupling features emerge when two or more systems are
taken under consideration \cite{Marwan,Frenzel,jie14}. Every one has heard of macroscopic coupling
in nature, e.g. in waves, interfaces, in modern life
such as social, economical $\&$ political issues \cite{CSE,EPL,Map}, coupling phenomena
in condensed matter physics \cite{elis13,Jamali} and in the context of
neuroscience \cite{wang14} are just some examples in this regard.
The most important aspect of coupling which motivated this work is
assessing the criticality or in other words the uncertainty born out
of coupled systems. Such coupling exists between companies, stock
markets, surfaces $\&$ interfaces, stochastic fields and, etc.

The multifractal formalism provides almost adequate tools for studying
the scaling relations and properties of objects possessing a fractal
geometry and/or generalized multifractal exponents
\cite{man971,man973,Bacry01,Bacry,cdfa}. Once, the
infinitely divisible cascades \cite{Castaing1,Chaboud,Muzy1} in
hydrodynamic turbulence was confirmed to be statistically
scale-invariant, one important motivation for multifractal
formulation to become prominent was established \cite{Nov90,Fris95}.
Multifractal models have been implemented in various fields of
sciences ranging from biology, geology, social to finance, e.g. Earth \& Solar winds
\cite{Sorrise,Telesca,hajian10}, foreign exchange rates
\cite{Ghas96}, stock index \cite{Kiy06,Sadegh}, %art \cite{art1,art2}, 
human heartbeat
fluctuations \cite{Kiy004,Kiy005}, well-log data \cite{log} seismic time series
\cite{Tabar09,Shadkhoo,Telesca0,Telesca1}, and sol-gel transitions
\cite{Shayeganfar1,Shayeganfar2,Shayeganfar3}.

A breakthrough in the applications of multifractal models took place when the relation between turbulence and finance in the
context of multiplicative random cascades was established \cite{Muzy1,Ghas96,Arneodo1}.
Note that their manifestation was based on their results that the
velocity increment fluctuations in turbulence and financial returns
are proportional through a stochastic factor. This factor only
depends on the scale ratio of the processes. In a further study, by
developing this theory, it has been shown that  a continuous random
walk model can posses multifractal properties due to the existence
of a correlation in the logarithm of the stochastic variances
\cite{Muzy1,Bacry}. It is worth noting that a strong log-normal
deviation from the normal case leads to a robust state of
multifractality or in other words a critical state in the underlying
system. This is due to the fact that, since the distribution
function has a fat tail, the occurrence of low frequent events gets
to be more probable compared to their corresponding in the normal
distribution. This is why and how the criticality enters the system.

In the present study we show how the individual uncertainty or criticality of each system affects their coupled
criticality. To this end, we study the criticality in coupled systems by implementing the bivariate multifractal random walk
method. However, it could be instructive to read the applications of this method in other disciplines \cite{Xiao,Lovs,Caren,Fauth,Lude}.

This paper is organized as follows: In section II method for
analysis will be explained. Data description and implementation of
method are explained in section III. Section IV is devoted to
summary and conclusions.

%%%%%%%%%%%%%%%%%%%%%%%%%%%%%%%%%%%%%%%%%%%%%%%%%%%%%%%%%%%%%%%%%%%%%%%%%%
\section{Model and analysis}
Consider a stochastic process represented by $x(t)$, which may be a
function of both space and time. More complementary explanations can be found in
\cite{Bacry,Brachet,Sornette}. Here we assume that time is a
dynamical parameter, therefore $x(t)$ is only taken to be time
dependant. The increment of $x(t)$ at time lag $\ell$ is defined as
$\Delta_{\ell} x(t)\equiv x(t+\ell)-x(t)$. According to the
cascading approach, the increment of fluctuations
$\Delta_{\ell}x(t)\equiv x(t+\ell)-x(t)$, at scales $\ell$ and $\eta
\times \ell$ satisfy the following relation
\begin{equation}\label{eq1}
\Delta_{(\eta\times \ell)} x(t)={\mathcal W}_{\eta} \Delta_{\ell}
x(t), \quad \forall \ \ell,\eta >0,
\end{equation}
where ${\mathcal W}_{\eta}$ is a stochastic variable \cite{Castaing1,Arneodo2}. We assume that the cascading process
starts from a large scale, $L$, where by starting the iterating process would eventually tend to
small scales ($\ell<L$). For a multiplicative cascading process which
starts from a large scale, $L$, tending to small scales, $\ell$, implementation of the multifractal random
walk approach enables us to rewrite the increment of fluctuation as $\Delta_{\ell}x(t)\equiv \xi_{\ell}(t) e^{\omega_{\ell}(t)}$, in which $\xi_{\ell}(t)$ and $\omega_{\ell}(t)$ are independent of each other and have Gaussian distributions with zero means.
The corresponding variances are denoted by $\sigma ^2(\ell)$ and $\lambda ^2(\ell)$ for $\xi_{\ell}(t)$ and $\omega_{\ell}(t)$,
respectively \cite{Bacry}. In this approach, a non-Gaussian probability density function (PDF) with fat tails is expressed by \cite{Castaing1}
\begin{equation}\label{eq2}
{\mathcal P}_{\ell} (\Delta_{\ell}x)=\int G_\ell (\ln\sigma(\ell))
\frac{1}{\sigma(\ell)} F_{\ell} \left(\frac{\Delta_{\ell}
x}{\sigma(\ell)}\right) d\ln \sigma(\ell),
\end{equation}
where we have
\begin{eqnarray}\label{gpdf}
%\begin{cases}
G_{\ell}(\ln \sigma(\ell))=\frac {1} {\sqrt{2\pi} \lambda(\ell)}\exp\left(-\frac{\ln ^{2}\sigma(\ell)} {2\lambda ^{2}(\ell)}\right), \\
F_{\ell} \left(\frac {\Delta _{\ell}x} {\sigma(\ell)}\right)=\frac
{1}{\sqrt{2\pi}} \exp\left(-\frac {\Delta_{\ell} x^{2}}
{2\sigma^{2}(\ell)}\right).
%\end{cases}
\end{eqnarray}
Simply, one can show that in the limit where $\lambda ^2(\ell)$
tends to zero, ${\mathcal P}_{\ell}(\Delta_{\ell}x)$ converges to a
Gaussian function. Increasing the parameter $\lambda ^{2}(\ell)$
quantifies the efficiency of the non-Gaussianity, where the tail of
the profile starts to fatten. Therefore, a large value of $\lambda
^{2}(\ell)$ indicates a high probability of finding large
fluctuations in a data set. This statement is a reminiscence of
criticality in the system as pointed out in \cite{Kiy06}. It is
worth noting the long-range correlation and/or non-Gaussianinty are
the sources of multifractality nature of a stochastic field. In this
paper we concentrate on the shape of the probability density function which acts a
source for multifractality characterized by $\lambda^{2}(\ell)$ \cite{bunde02,Graho}.

Since it is possible that two neighbouring sites interact with each
other, the verification of their correlation would be of interest.
This brings up the idea that the coupled behaviour may correspond to
the behaviour of each individual system together with their cross
correlation. In this line Muzy et al. considered the cross  correlation between processes by generalizing
the multifractal random walk approach based on the log-normal
cascade model \cite{Muzy2}. This generalization introduces the
multivariate multifractal model which describes the scale invariance
of the joint statistical properties. Suppose $\textbf{x}(t)\equiv \{x_{1}(t),x_{2}(t)\}$ is represented
as a bivariate process, the bivariate version of $BiMRW$ for the
increment $\Delta_{\ell}\textbf{x}(t)=\textbf{x}(t+\ell)-\textbf{x}(t)$ reads
as \cite{Muzy2}
\begin{equation}\label{eq3}
\Delta_{\ell}\textbf{x}(t)=\left(\xi_{\ell}^{(1)}(t)e^{\omega_{\ell}^{(1)}(t)},\xi_{\ell}^{(2)}(t)
e^{\omega_{\ell}^{(2)}(t)}\right),
\end{equation}
where for each increment at the time lag $\ell$ there exists a
separate $\xi$ and $\omega$. Note that the bivariate processes
$[\xi_{\ell}^{(1)},\xi_{\ell}^{(2)}]$ and
$[\omega_{\ell}^{(1)},\omega_{\ell}^{(2)}]$ are independent of one
another, both having a joint Gaussian distributions with a zero
mean. The covariance matrices $\mathbf{\Sigma}_{\ell}$ and
$\mathbf{\Lambda}_{\ell}$ respectively become
\begin{eqnarray}\label{coe1}
\mathbf{\Sigma}_{\ell}&\equiv&\left(
               \begin{array}{cc}
                 \Sigma_{\ell}^{(11)} & \Sigma_{\ell}^{(12)}\\
                 \Sigma_{\ell}^{(21)} & \Sigma_{\ell}^{(22)} \\
               \end{array}
             \right) \nonumber\\
             &=&\left(
               \begin{array}{cc}
                 \langle \xi_{\ell}^{(1)}(t) \xi_{\ell}^{(1)}(t)\rangle &  \langle \xi_{\ell}^{(1)}(t) \xi_{\ell}^{(2)}(t)\rangle \\
                  \langle \xi_{\ell}^{(2)}(t) \xi_{\ell}^{(1)}(t)\rangle &  \langle \xi_{\ell}^{(2)}(t)\xi_{\ell}^{(2)}(t)\rangle\nonumber\\
               \end{array}
             \right),\\
             \mathbf{\Lambda}_{\ell}&\equiv&\left(
                                                         \begin{array}{cc}
                                                           \Lambda_{\ell}^{(11)} & \Lambda_{\ell}^{(12)} \\
                                                           \Lambda_{\ell}^{(21)} & \Lambda_{\ell}^{(22)} \\
                                                         \end{array}
                                                       \right)\nonumber\\
                                                       &=&\left(
                                                         \begin{array}{cc}
                                                          \langle \omega_{\ell}^{(1)}(t) \omega_{\ell}^{(1)}(t)\rangle &  \langle \omega_{\ell}^{(1)}(t) \omega_{\ell}^{(2)}(t)\rangle \\
                  \langle \omega_{\ell}^{(2)}(t) \omega_{\ell}^{(1)}(t)\rangle &  \langle \omega_{\ell}^{(2)}(t)\omega_{\ell}^{(2)}(t)\rangle\nonumber\\
                                                         \end{array}
                                                       \right).\\
\end{eqnarray}
The four diagonal elements of the presented matrixes, namely
$\Sigma_{\ell}^{(11)}\equiv\sigma_1^2(\ell)\ ,\
\Sigma_{\ell}^{(22)}\equiv\sigma_2^2(\ell)$,
$\Lambda_{\ell}^{(11)}\equiv\lambda_1^2(\ell)$, and $\Lambda_{\ell}^{(22)}\equiv\lambda_2^2(\ell)$ are defined for the
two individual processes $1$ and $2$. In addition, the symmetry
property of these matrices implies the following equalities;
$\Sigma_{\ell}^{(12)}=\Sigma_{\ell}^{(21)}\equiv\Sigma_{\ell}
\sigma_1(\ell)\sigma_2(\ell)$ and
$\Lambda_{\ell}^{(12)}=\Lambda_{\ell}^{(21)}\equiv\Lambda_{\ell}
\lambda_1(\ell)\lambda_2(\ell)$. Usually,  $\mathbf{\Sigma}_{\ell}$
is called the "{\it Markowitz matrix}" which shows the variance
and correlation of $\xi$'s. $\mathbf{\Lambda}_{\ell}$ is the
"{\it multifractal matrix}" which quantifies the non-linearity of $\omega$'s \cite{Muzy2,marko}. %However
%their importance for us is due their correspondence to the intensity of
%the cross-correlation.
In this framework, the shape of joint-PDF would be
\begin{eqnarray}\label{eq5}
&&{\mathcal P}_{\ell}
\left(\Delta_{\ell}x_{1},\Delta_{\ell}x_{2}\right)=
\int d(\ln \sigma_{1}(\ell)) \int d(\ln \sigma_{2}(\ell))\nonumber\\
&&G_{\ell}\left(\ln \sigma_{1}(\ell) , \ln \sigma_{2}(\ell)\right)
\frac{1} {\sigma_{1}(\ell)\sigma_{2}(\ell)} F_{\ell} \left(\frac
{\Delta_{\ell} x_{1}} {\sigma_{1}(\ell)},\frac {\Delta_{\ell} x_{2}}
{\sigma_{2}(\ell)}\right),\nonumber\\
\end{eqnarray}
where $G_{\ell}(\ln \sigma_{1}(\ell),\ln \sigma_{2}(\ell))$ and
$F_{\ell}\left(\frac{\Delta_{\ell}
x_1}{\sigma_1(\ell)},\frac{\Delta_{\ell}
x_2}{\sigma_2(\ell)}\right)$ are the probability density functions
of the bivariate processes
$(\omega_{\ell}^{(1)},\omega_{\ell}^{(2)})$ and
$(\xi_{\ell}^{(1)},\xi_{\ell}^{(2)})$, respectively. By taking into
account the cross correlation between the two systems under
consideration, their joint-probability density function in the
integral of would be
\begin{widetext}
\begin{eqnarray}
G_{\ell}\left(\ln \sigma_{1}(\ell) , \ln
\sigma_{2}(\ell)\right)=\frac {1}{2\pi\lambda _{1}(\ell) \lambda
_{2}(\ell) \sqrt{(1-\Lambda_{\ell}
^{2})}}\times\,\,\,\,\,\,\,\,\,\,\,\,\,\,\,\,\,\,\,\,\,\,\,\,\,\,\,\,\,\,\,\,\,\,\,\,\,\,\,\,\,\,\,\,\,\,\,\,\,\,\,\,\,\,\,\,\,\,\,\,\,\,\,\,\,\nonumber\\
\mathrm{exp} \left(-\frac {1} {2(1-\Lambda_{\ell} ^{2})}
\left[\left(\frac {\ln ^{2}\sigma_{1}(\ell)} {\lambda _{1}^{2}(\ell)
}\right)+\left(\frac {\ln ^{2}\sigma_{2}(\ell)} {\lambda
_{2}^{2}(\ell)}\right)- 2\Lambda_{\ell}
 \left(\frac{\ln{\sigma_{1}(\ell)}}{\lambda _{1}(\ell)}\right)\left(\frac{\ln{\sigma_{2}(\ell)}}{\lambda _{2}(\ell)} \right)\right]\right), \nonumber
\end{eqnarray}
and
\begin{eqnarray}
 F_{\ell}\left(\frac {\Delta_{\ell} x_{1}} {\sigma_{1}(\ell)},\frac{\Delta_{\ell} x_{2}} {\sigma_{2}(\ell)}\right)=\frac {1} {2\pi
\sqrt{(1-\Sigma_{\ell} ^{2})}}\times\,\,\,\,\,\,\,\,\,\,\,\,\,\,\,\,\,\,\,\,\,\,\,\,\,\,\,\,\,\,\,\,\,\,\,\,\,\,\,\,\,\,\,\,\,\,\,\,\,\,\,\,\,\,\,\,\,\,\,\,\,\,\,\,\,\,\,\,\,\,\,\,\,\,\,\,\,\,\nonumber\\
\mathrm{exp} \left(-\frac {1} {2(1-\Sigma_{\ell} ^{2})} \left[
\left(\frac {\Delta_{\ell} x_{1} ^{2}}
{\sigma_{1}^{2}(\ell)}\right)+\left(\frac {\Delta_{\ell} x_{2} ^{2}}
{\sigma_{2} ^{2}(\ell)}\right)- 2\Sigma_{\ell}
\left(\frac{\Delta_{\ell}
x_{1}}{\sigma_{1}(\ell)}\right)\left(\frac{\Delta_{\ell} x_{2}}
{\sigma_{2}(\ell)}\right) \right] \right).
\end{eqnarray}
\end{widetext}
It turns out that if the cross correlation coefficients
$\Lambda_{\ell}$ and $\Sigma_{\ell}$  are to be zero, consequently
$G_{\ell}$ and $F_{\ell}$ would simply be the multiplication of the
two independent processes.
This result confirms that any deviation from the product of these
two independent processes would lead to the coupling of the two
processes. According to the covariance matrix defined in Eq.
(\ref{coe1}), the parameter $\Lambda_{\ell}$ controls the strength
of the joint multifractality for the two processes \cite{Muzy2}.

In order to estimate the parameter $\Lambda_{\ell}$ at scale $\ell$, we
use the Bayesian statistics \cite{gd03,sivv06}. Consider an original data set and a theoretical model that is able to create the total of N non-Gaussian data sets with corresponding $(N)$ theoretical PDFs. For each point of the original PDF, there exist $N$ equivalent points extracted from the theoretical PDFs, where due to the central limit theory their PDF is Gaussian. In this line we write the corresponding likelihood function \cite{udo11}, $\mathcal{L}$, in the form of a multivariate Gaussian function is
%\begin{widetext}
\begin{eqnarray}\label{eq771}
&&{\mathcal{L}}({\mathcal{P}}_{\rm data}(\textbf{y},\ell)|{\mathcal{P}}_{\rm theory}(\textbf{y};\mathbf{\Sigma}_{\ell},\mathbf{\Lambda}_{\ell}))\\\nonumber
&&=\frac{\sqrt{{\mathcal
{\rm Det}\{{\mathcal F}\}}}}{(2\pi)^{N/2}}\exp \left( -\frac{\Delta^{T}.{\mathcal F}.\Delta}{2}\right)
\end{eqnarray}
%\end{widetext}
where $\Delta\equiv {\mathcal{P}}_{\rm
data}(\textbf{y},\ell)-{\mathcal {P}}_{\rm
theory}(\textbf{y};\mathbf{\Lambda}_{\ell},\mathbf{\Sigma}_{\ell})$ is a column vector, ${\mathcal{F}}$ is the fisher information matrix which is determined according to
${\mathcal{F}}^{-1}=\langle \Delta({\textbf{y}}) \Delta({\textbf{y}}') \rangle $. ${\textbf{y}}$ is an independent parameter and
$(\mathbf{\Lambda}_{\ell},\mathbf{\Sigma}_{\ell})$ is a model free parameter. ${\mathcal {P}}_{\rm data}(\textbf{y},\ell)$ is computed directly from the data sets, and ${\mathcal {P}}_{\rm
theory}(\textbf{y};\mathbf{\Lambda}_{\ell},\mathbf{\Sigma}_{\ell})$ is estimated from Eq.
(\ref{eq5}). Since there is no reason to have cross correlation between $\Delta({\textbf{y}})$'s for
different ${\textbf{y}}$'s the fisher matrix would be diagonal (if there are non-zero cross correlation in covariance matrix then one can use proper similarity transformation to diagonalize this Hermitian matrix). The best fit for all scales which has the maximum likelihood with the original PDF, is found using the $\chi^{2}$ test. This is where $\chi^{2}$ obtains its global minimum
\begin{equation}\label{eq77}
\chi^{2}(\mathbf{\Lambda}_{\ell};\mathbf{\Sigma}_{\ell})=\sum_{\textbf{y}}\frac{[{\mathcal{P}}_{\rm
data}(\textbf{y},\ell)-{\mathcal {P}}_{\rm
theory}(\textbf{y};\mathbf{\Lambda}_{\ell};\mathbf{\Sigma}_{\ell})]^{2}}{\sigma^{2}_{\rm
data}(\textbf{y},\ell)+\sigma^{2}_{\rm theory}(\textbf{y};
\mathbf{\Lambda}_{\ell};\mathbf{\Sigma}_{\ell})},
\end{equation}
where $\sigma^{2}_{\rm data}(\textbf{y},\ell)$ and $\sigma^{2}_{\rm
theory}(\textbf{y};\mathbf{\Lambda}_{\ell},\mathbf{\Sigma}_{\ell})$
are the mean standard deviation of ${\mathcal {P}}_{\rm
data}(\textbf{y})$ and ${\mathcal {P}}_{\rm
theory}(\textbf{y};\mathbf{\Lambda}_{\ell},\mathbf{\Sigma}_{\ell})$,
respectively. By marginalizing over nuisance free parameter, $\mathbf{\Sigma_{\ell}}$,  we obtain
\begin{equation}\label{eq7}
\chi^{2}(\mathbf{\Lambda}_{\ell})=\sum_{\textbf{y}}\int
d\mathbf{\Sigma}_{\ell} \left(\frac{[{\mathcal{P}}_{\rm
data}(\textbf{y},\ell)-{\mathcal {P}}_{\rm
theory}(\textbf{y};\mathbf{\Lambda}_{\ell},\mathbf{\Sigma}_{\ell})]^{2}}{\sigma^{2}_{\rm
data}(\textbf{y},\ell)+\sigma^{2}_{\rm theory}(\textbf{y};
\mathbf{\Lambda}_{\ell},\mathbf{\Sigma}_{\ell})}\right).
\end{equation}
The best fit values for the non-Gaussian parameters
$\mathbf{\Lambda}_{\ell}$ are determined systematically by searching
in the landscape of marginalized chi-square.

%@@@@@@@@@@@@@@@@@@@@@@@@@@@@@@@@@@@@@@@@@@@@@@@@@@@@@@@@@@@@@@@@@@@@@@@@@@@@@@@@@@@@@@@@@
\begin{figure*}[ht]
%\begin{figure}[t]
\centering
\includegraphics[scale=0.9]{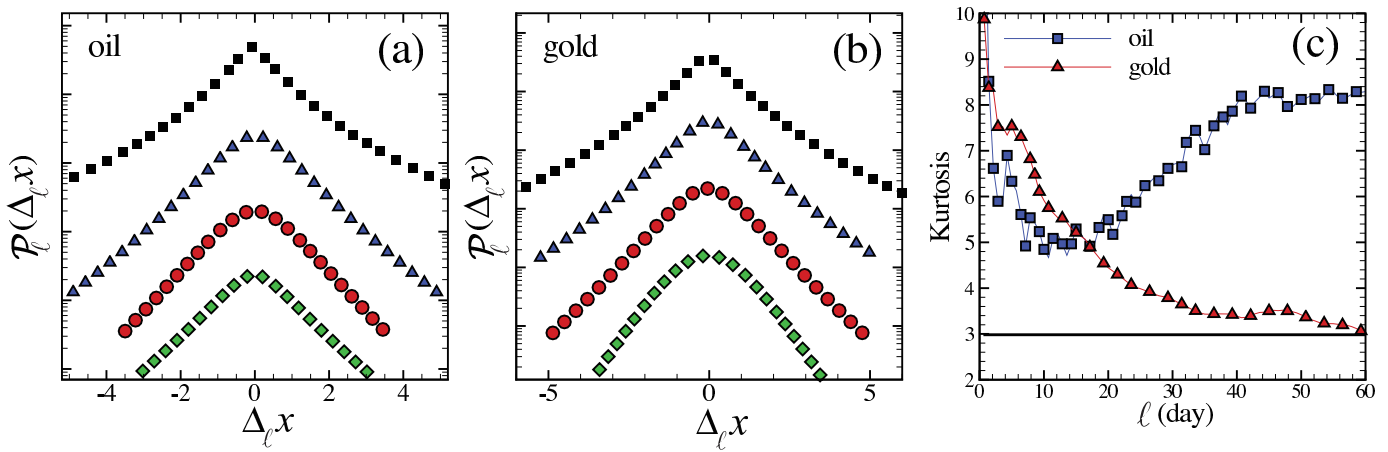}
\caption{ \label{fig2} Panels (a) and (b) indicate probability
density function of oil and gold markets from top to bottom for
daily (square symbol), weekly (delta symbol), monthly (circle
symbol), and seasonally (diamond symbol) time scales. Panel (c)
shows Kurtosis of the markets versus time lag, $\ell$.}
%\end{figure}
\end{figure*}
%@@@@@@@@@@@@@@@@@@@@@@@@@@@@@@@@@@@@@@@@@@@@@@@@@@@@@@@@@@@@@@@@@@@@@@@@@@@@@@@@@@@@@@@@@

The cross correlation function of the processes is obtained by
\begin{equation}
 C_{\ell}^{\rm joint} (\tau)\equiv \langle[\bar{\omega}_{\ell}^{(1)} (i) -\langle\bar{\omega}_{\ell}^{(1)} \rangle][\bar{\omega}_{\ell}^{(2)} (i+\tau)-
\langle \bar{\omega}_{\ell}^{(2)} \rangle]\rangle.
\end{equation}
As stated earlier, the upper indices $"1"$ and $"2"$ refer to the processes $(1)$ and $(2)$ respectively, and $\tau$ represents the time lag where $\ell<\tau$. Due to the fact that the vector process is stationary, the cross correlation is only dependant on $\tau$. The parameter $\bar{\omega}^{(\diamond)}_{\ell} (i)$ is the local variance defined
by
\begin{equation}
\bar{\omega}^{(\diamond)}_{\ell}(i) =\frac{1} {2}
\ln\sigma_{(\diamond)}^{2} (\ell,i),
\end{equation}
where its magnitude is
\begin{equation}
\sigma_{(\diamond)}^{2} (\ell;i)=\frac{1} {\ell}
\sum_{j=1+(i-1)\ell} ^{i\ell} \Delta_{\ell} x_{\diamond}^2(j).
\end{equation}
The symbol $(\diamond)$ should be replaced by $(1)$ and $(2)$ for the first and second data sets \cite{Kiy06,Kiy004,Kiy005,Arneodo3}.

%%%%%%%%%%%%%%%%%%%%%%%%%%%%%%%%%%%%%%%%%%%%%%%%%%%%%%%%%%%%%%%%%%%

\section{Applications of BiMRW}

In this section, we apply our method to the reciprocal
effects of oil and gold markets which have been recorded on daily
basis at the time interval from $1995$ to $2012$ \cite{site2}.

As discussed in detail earlier in the text, in {\it BiMRW}, the joint
multifractal parameter, $\Lambda_{\ell}$, describes the coupling of
large fluctuations of two processes \cite{Muzy2}. As a matter of
fact a large value of $\Lambda_{\ell}$ refers to a robust joint
multifractality which results in a coupled criticality or
uncertainty state in the system. The scaling parameter,
$\Lambda_{\ell}$, which plays an important role in
$G_{\ell}\left(\ln\sigma_1(\ell),\ln\sigma_2(\ell)\right)$, can be written as \cite{Muzy2}:%,Newland}
\begin{equation}\label{eq8}
\Lambda_{\ell}=\frac{\Lambda_{\ell}^{(12)}}{\lambda_{1}(\ell)
\lambda_{2}(\ell)}=\frac {\langle\ln \sigma_{1}(\ell) \ln
\sigma_{2}(\ell)\rangle}{\lambda_{1}(\ell) \lambda_{2}(\ell)},
\end{equation}
where $\langle \cdots \rangle$ denotes the ensemble averaging on all
windows with size $\ell$. It could be deduced from Eq. (\ref{eq8}) that the
scaling parameter, $\Lambda_{\ell}$, is affected by two parameters
namely, the non-Gaussian parameters $\lambda$'s, and the cross
correlation of the stochastic variances $\langle\ln
\sigma_{1}(\ell)) \ln\sigma_{2}(\ell)\rangle$. Note that since
$\Lambda_{\ell}$ depends on the scale $\ell$, the scale in which
$\Lambda_{\ell}$ ascends must be carefully investigated. Actually, a
large $\Lambda_{\ell}$ implies the emergence of coupled criticality
or uncertainty.
The various situations that $\Lambda_{\ell}$ probably comes up are as follows:\\

 - if two underlying systems are uncorrelated, due to independency of associated individual uncertainty which are quantified by $\lambda$'s, consequently the systems will not experience any coupling. Hence there is no uncertainty between them.\\

 - in case of correlated systems, the coupled criticality can emerge when at least one of underlying systems possess a Gaussian distribution. In this situation, the $\lambda$'s tend to zero. Such condition is a benchmark of resonance. However, in this normal state, the weak cross correlation of large fluctuations results in decreasing $\Lambda_{\ell}$.\\

 - if two underlying systems are correlated, but both of them are in their high criticality states, by increasing the $\lambda_{\ell}$'s, the coupled uncertainty decreases.\\

Regarding these explanations we will asses how the coupled
uncertainty exists and examine whether this coupled uncertainty is
generally directed or not. In principle, we expect that a system associated with a more uncertainty state with large $\lambda$'s, would lead to the occurrence of large fluctuations causing an uncertainty state in the neighbouring system. This implies that the two systems are correlated, therefore, the coupled uncertainty becomes large. This means that $\Lambda_{\ell}$ also increases. Subsequently, one can deduce that $\Lambda_{\ell}$ behaves similar to the conditional uncertainty of a system that is more normal and impressed by the other system. In contrast, if the system that is in the normal state causes large fluctuations in the neighbouring system which is in a higher uncertainty state, the cross correlation of large fluctuations decreases, resulting in decreasing of $\Lambda_{\ell}$.\\

\begin{figure}[t]
\centering
\includegraphics[scale=0.55]{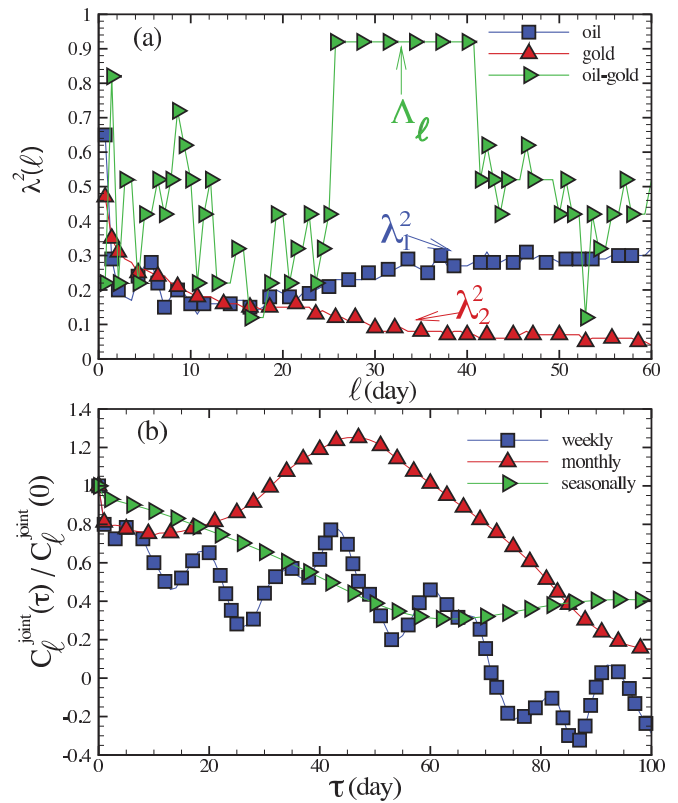}
\caption{ The scale dependence of the non-Gaussian parameter,
$\lambda ^{2}(\ell)$, for oil and gold markets recorded daily at the
time interval between $1995$ to $2012$ has been shown in panel (a).
The coupling multifractal parameter $\Lambda_{\ell}$ for the joint
markets has been also plotted versus scale, $\ell$. Panel (b)
illustrates the magnitude of cross correlation function of oil and
gold markets, $C_{\ell} ^{\rm joint}(\tau)/C_{\ell} ^{\rm joint}(0)$
versus time lag, $\tau$, for weekly (square symbol), monthly
(triangle symbol) and seasonally (right triangle symbol)
scales.}\label{fig1}
\end{figure}

The corresponding stochastic parameters for further investigation
are daily records of $x(t)$ for oil and gold markets \cite{site2}.
The analysis is based on the stationary increment fluctuations of
the markets. According to Eq. (\ref{eq3}), the bivariate model for
underlying data sets have been determined. Probability density
function based on cascading approach have been illustrated in Fig.
\ref{fig2}. This plots confirm that a non-Gaussian behaviour at
scales smaller than a month exists for both series. Gold market has
a Gaussian probability density function at larger scales. This
statement is proved by panel (c) of Fig. \ref{fig2}, where the
kurtosis of the markets has been plotted. One can also deduce that
the kurtosis of the markets at scales less than a month is greater
than $3$. For scales greater than a month, the kurtosis for the gold
market reaches to $3$. Thus, one would expect a Gaussian
distribution for the gold market at scales much larger than a month.
To get more reliable results, we have also computed non-Gaussian
parameter $\lambda_{\diamond}^2(\ell)$ according to the likelihood
statistics explained in the previous section for both data sets and
indicated in Fig. \ref{fig1}. As shown in panel (a) of Fig.
\ref{fig1}, the value of $\lambda^{2}_{1}(\ell)$ for the oil market
is high at all scales especially  at large scales, however, for the
gold market the large value of $\lambda^{2}_2(\ell)$ turns up at
small scales. Thus, the oil market resembles a market with a
continuous criticality compared to the gold market which shows
criticality only at small scales.
%One can deduce from panel (a) of Fig. \ref{fig1}, non-Gaussian parameter has large value at small scale represented highly non-Gaussianity in underlying data and by increasing the scale $\lambda_1^2(\ell)$ for oil decreases %but around $\ell \gtrsim$ month a small increasing was found.
This behaviour is in agreement with Fig. \ref{fig2} (a). In addition, the
joint multifractal parameter, $\Lambda_{\ell}$, has been plotted
versus scales $\ell$ in Fig. \ref{fig1}. It could be noticed that
$\Lambda_{\ell}$ rises during the time interval $[25-40]$ days. A large
value of $\Lambda_{\ell}$ reveals a state with a strong coupled
criticality that is a confirmation for the existence of resonance in
the market. This is due two following reasons; the first comes from
the normal behaviour of the gold market
($\lambda^2_{\rm{gold}}=\lambda^2_{2}\rightarrow0$, indicated in
Fig. \ref{fig1} (a)). The second is because of a powerful coupling of
large fluctuations of the oil and gold markets. This is due to a
long-range cross correlation of the stochastic magnitudes at a
monthly scale, see Fig. \ref{fig1} (b). For time scales before and
after the powerful coupling the scenario is different. For time
scales before the strong coupling or in other words time scales
smaller than a month, the joint multifractal parameter,
$\Lambda_{\ell}$, shows a decreasing in the market. Indeed the
existence of an individual criticality state or in other words large
value of $\lambda^2_{\diamond}(\ell)$'s for each market accompanied
by a weak magnitude cross correlation for a weekly scale (Fig.
\ref{fig1} (b)) results in the decreasing of the coupled criticality
(reducing $\Lambda_{\ell}$). On the other hand, as shown in Fig.
\ref{fig1} (b),  at large scales or scales after the strong
coupling, the oil market with large $\lambda^2_1(\ell)$ has a
short-range magnitude cross correlation with the gold market for a
seasonal scale. This issue again causes a decrease of criticality in
coupled markets.
%@@@@@@@@@@@@@@@@@@@@@@@@@@@@@@@@@@@@@@@@@@@@@@@@@@@@@@@@@@@@@@@@@@@@@@@@@@@@@@@@@@@@@@@@@
%@@@@@@@@@@@@@@@@@@@@@@@@@@@@@@@@@@@@@@@@@@@@@@@@@@@@@@@@@@@@@@@@@@@@@@@@@@@@@@@@@@@@@@@@@
\begin{figure*}[ht]
%\begin{figure}[t]
\includegraphics[scale=0.9]{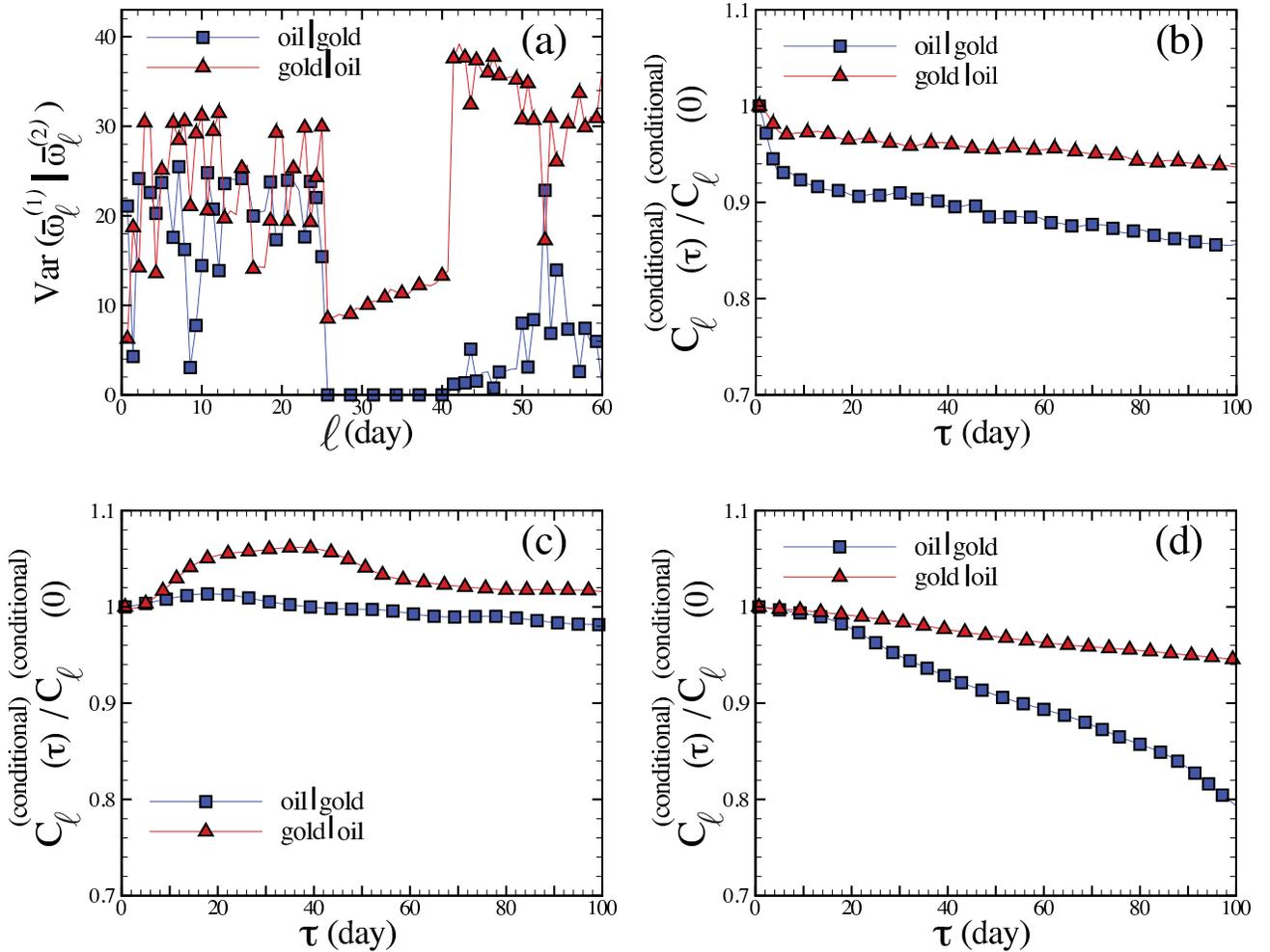}
\centering \caption{Panel (a): Variance of the conditional
distribution of the local magnitudes versus scale, $\ell$, for oil
and gold markets. Panels (b), (c) and (d) indicate the conditional
magnitude cross correlation function of oil and gold markets,
$C_{\ell}^{\rm{conditional}}(\tau)/C_{\ell}^{\rm{conditional}}(0)$,
as a function of time lag $\tau$,  for weekly, monthly and
seasonally scales, respectively.}\label{fig3}
%\end{figure}
\end{figure*}
%@@@@@@@@@@@@@@@@@@@@@@@@@@@@@@@@@@@@@@@@@@@@@@@@@@@@@@@@@@@@@@@@@@@@@@@@@@@@@@@@@@@@@@@@@
A question that arises here is whether the occurrence of an
uncertainty state in one system produced by the other system is a
directed phenomenon or not. In order to answer this question, we
must first show how the occurrence of large fluctuations and a high
criticality in one system are affected by large fluctuations in the
other system. To this end, we should evaluate the variance of the
conditional distribution for the stochastic local magnitudes. The
variance of conditional distribution of the local magnitudes is
defined by
\begin{equation}\label{eq11}
{\rm{Var}}\left(\bar{\omega}_{\ell}^{(1)}
|\bar{\omega}_{\ell}^{(2)}\right)\equiv\sum_{i}
\left[\bar{\omega}_{\ell}^{(1)} (i)-\langle
\bar{\omega}_{\ell}^{(1)} \rangle\right]^{2} {\mathcal
P}\left(\bar{\omega}_{\ell}^{(1)} (i)|\bar{\omega}_{\ell}^{(2)}
(i)\right),
\end{equation}
where the conditional distribution function ${\mathcal
P}(\bar{\omega}_{\ell}^{(1)} (i)|\bar{\omega}_{\ell}^{(2)} (i))$,
can be read as
\begin{equation}
{\mathcal P}\left(\bar{\omega}_{\ell}^{(1)}
(i)|\bar{\omega}_{\ell}^{(2)} (i)\right)=\frac{{\mathcal
P}\left(\bar{\omega}_{\ell}^{(1)} (i),\bar{\omega}_{\ell}^{(2)}
(i)\right)}{{\mathcal P}\left(\bar{\omega}_{\ell}^{(2)} (i)\right)},
\end{equation}
where ${\mathcal P}(\bar{\omega}_{\ell}^{(1)}
(i),\bar{\omega}_{\ell}^{(2)} (i))$, is the joint distribution
function of the stochastic local magnitude,
$\bar{\omega}_{\ell}^{(\diamond)}$, for the two underlying
processes. The large conditional variance is due to the occurrence
of large fluctuations in a system through the fluctuations in the
other system. Nonetheless, a system with higher criticality
condition (large $\lambda^2_{\diamond}(\ell)$) would result in the
occurrence of large fluctuations accompanied by a high conditional
criticality in the other system. This phenomenon is due to a strong
cross correlation of large fluctuations between the two systems.
Now, we introduce the concept of directed coupled criticality.
Strictly speaking this concept means that if a coupling between two
systems is considered as non-balanced, or e.g. in social life as a
one sided relation such as a one sided love, one of the people
(systems) would be in an uncertainty (criticality) state with the
other while this is not true (or an option) the other way round. In
order to investigate this directed coupled uncertainty, the
conditional cross correlation function of processes $(1)$ and $(2)$
at time scale $\ell$ versus time lag $\tau$ ($\ell<\tau$) must be
defined. Therefore, the conditional cross correlation is:
\begin{eqnarray}\label{crosscondi}
&&C^{(\rm{conditional})}_{\ell}(\tau)\equiv\nonumber\\
&&\sum_{i} \left[\bar{\omega}_{\ell}^{(1)} (i)-\langle
\bar{\omega}_{\ell}^{(1)}
\rangle\right]\left[\bar{\omega}_{\ell}^{(2)} (i+\tau)-\langle
\bar{\omega}_{\ell}^{(2)}
\rangle\right]\nonumber\\
&& \times {\mathcal P}\left(\bar{\omega}_{\ell}^{(1)}
(i)|\bar{\omega}_{\ell}^{(2)} (i)\right).
\end{eqnarray}
Fig. \ref{fig3}(a) shows the conditional variance of the stochastic
magnitudes of the markets corresponding to the definition expressed
by Eq. (\ref{eq11}). The conditional magnitude variance of the gold
market is more than the oil market at almost all scales. This means
that the oil market with a large uncertainty or criticality produces
large fluctuations in the gold market. This originates the fact that
the conditional cross correlation of the gold market is long-range
at all scales, see panels (b,c,d) of Fig. \ref{fig3}. The comparison
of Fig. \ref{fig3} (a) and Fig. \ref{fig1} (a) shows that the
behaviour of the conditional variance of the oil market,
${\rm{Var}}\left(\bar{\omega}_{\ell}^{({\rm oil})}
|\bar{\omega}_{\ell}^{({\rm gold})}\right)$ is almost opposite to
behaviour of $\Lambda_{\ell}$. It must point out that, the variance
of the oil market which is considered to have a high uncertainty, is
caused  by the gold market (which is a normal market), would result
in a weak cross correlation between the two markets. Hence, the
coupled uncertainty decreases, meaning that $\Lambda_{\ell}$
decreases. It is readily noticed by Fig. \ref{fig3} (a) that at the
resonant region, about a one month time-scale, the conditional
variance of the oil market tends to zero. This expresses that the
gold market which is normal at this time scale
($\lambda^2_{\rm{gold}}(\ell)\rightarrow0$) does not produce
noticeable changes in the oil market, thus, the conditional variance
of the oil market takes its minimum value.

It is instructive to provide another application in compliance with
the discussions of the present study. Due to the different reports,
it is known that inflation and unemployment posses an inverse
proportionality with each other. In other words, decreasing of one
of them causes to increase other \cite{Philip}. This has been a
trick for governments to control one by the other one. As of the
context of the present study the interest is to look at inflation
and unemployment from the view point of coupled criticality.

\begin{figure}[h]
\includegraphics[scale=0.52]{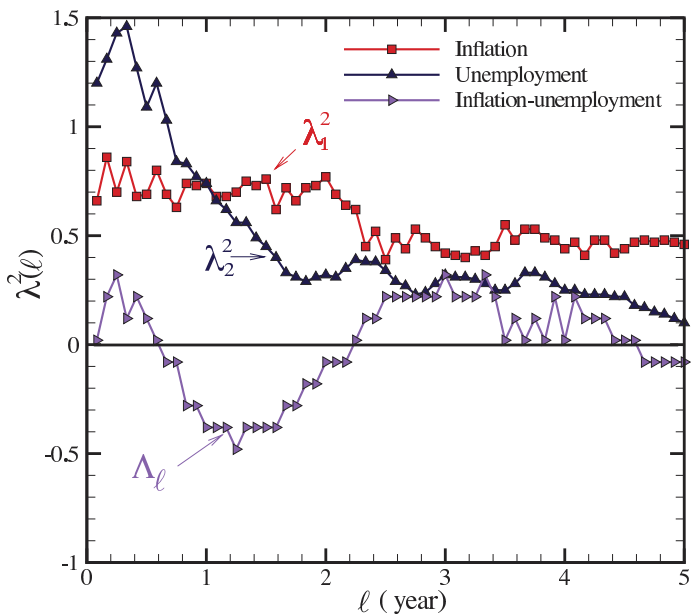}
\caption{The non-Gaussian parameter for inflation (square symbol)
rates and unemployment ( triangle symbol)  together with their joint
multifractal parameter $\Lambda_{\ell}$ (right triangle symbol). The
periods where $\Lambda_{\ell}$ is positive means that inflation and
unemployment rates are directly proportional to each other. From
around month $6$ to month $26$ where $\Lambda_{\ell}$ is negative,
inflation and unemployment rates are inversely proportional to each
other.} \label{fig5}
\end{figure}

By extracting data from the U.S. Inflation Calculator and the U.S.
Bureau of Labor Statistics \cite{site} we study the time series of
inflation and unemployment rates and compare their non-Gaussian
parameters, $\lambda^2_{\diamond}(\ell)$. It is shown in Fig.
\ref{fig5} that for small scales the non-Gaussianity of unemployment
rates is greater than for inflation rates. But towards larger scales
it is the inflation rate that possesses a greater non-Gaussian
parameter. But most interesting of all is their joint non-Gaussian
parameter ($\Lambda_{\ell}$) which indicates that in small scales
and large scales their behaviour is proportional to each other
($\Lambda_{\ell}>0$), while at the intermediate range, namely from
month $6$ to month $26$ their behaviour is inversely proportional to
each other $(\Lambda_{\ell}<0)$. This provides an insight on the
joint uncertainty or coupled criticality represented by
$\Lambda_{\ell}$ which is non-zero, regarding the long-range cross
correlation of large fluctuations between two systems.
%@@@@@@@@@@@@@@@@@@@@@@@@@@@@@
%@@@@@@@@@@@@@@@@@@@@@@@@@@@@
\section{Conclusions}
In this work, the bivariate multifractal random walk method has been
implemented in order to describe the criticality or uncertainty
emerged from the coupled multifractality of two systems in the
context of the joint log-normal cascade model. As coupling is a
scaling concept,  therefore we have indicated in this paper that
there are certain scales that possess strong couplings. The
parameter under investigation was the joint multifractal parameter,
$\Lambda_{\ell}$, illustrated at various scales, $\ell$. The
estimation of $\Lambda_{\ell}$ revealed its dependence on both the
non-Gaussian parameter, $\lambda^2_{\diamond}({\ell})$, of each
individual process, and the cross correlation of the stochastic
variances. It was also shown that the uncertainty of the coupled
systems is non-symmetric. This means that the occurrence of large
fluctuations in one of the two systems caused by the other system
may differ if it were to be from the other way round.

The value of the non-Gaussianinty parameter in the context of the present study ($\lambda_{\diamond}^2(\ell)$) decreases with $\ell$ for both of the series at intermediate scales which is around one month. This behaviour remains just for gold data while for oil we
found that after passing the intermediate scale, an increase for
$\lambda_{\rm {oil}}^2(\ell)$ takes place. As indicated in Fig.
\ref{fig1} (a), our results confirmed that there exists a powerful
coupling together with a high uncertainty between oil and gold
markets around a one month time-scale in a situation that the gold
market is tending to a normal state
($\lambda^2_{\rm{gold}}\rightarrow 0$). In order to show which market causes large fluctuations in the other market, variance of
the conditional distribution of the stochastic local magnitudes of
each data set have been evaluated. Our results corroborated that the
large fluctuations in the oil market give rise to the occurrence of
large fluctuations in the gold market forming a high conditional
variance of gold$\mid$oil. However, this is not established for the
conditional variance of oil$\mid$gold, as indicated in Fig.
\ref{fig3} (a), where the conditional variance of gold$\mid$oil is
greater than the conditional variance of oil$\mid$gold. The
conditional cross correlation (Eq. (\ref{crosscondi})) of
gold$\mid$oil is grater than for oil$\mid$gold at all time scale
(Fig. \ref{fig3} (b,c,d)). These results confirmed that the oil
market imposes large fluctuations on the gold market and it is
consistent with the results given by the conditional variance (Eq.
(\ref{eq11})).

The results obtained here confirm that when the systems are in their
normal situations and the cross correlations of large fluctuations
are long-range, the joint multifractality, $\Lambda_{\ell}$, is
strongest. This level was hereby chosen to be called the resonance
state. In the case where at least one of the systems is in a high
uncertainty state, $\Lambda_{\ell}$ decreases. This issue is caused
by the weak cross correlations between the systems. An important
finding of this article is deduced from Eq. (\ref{eq8}) which shows
that in order to control the uncertainty in a coupled system,
creating an uncertainty state in both of the two individual systems
in the presence of their cross correlation proves adequate for a
least uncertainty in the coupled system. This is a sign of
reciprocal or in other words depending fluctuations between two
systems. This would result in great joint changes in the two systems
involved.

The proposed method has been applied to two financial time series.
The conclusion is that there exists a robust coupling of the oil and
gold markets around one month time-scale in the presence of
long-range cross correlations between the two markets. However
before and after the one month time scale the systems behave
differently. In a sense that the reduction of coupled uncertainty is
due to the weak cross correlation of large changes between the two
systems. This proves the importance of the resonance region which
for the gold and oil markets is about one month. This may provide
opportunity for organizations that experience loss due to oil or
gold fluctuations to be able to plan ahead for preventing loss or
gaining benefit.

{\bf Acknowledgements:} S.M.S. Movahed and G.R. Jafari are thankful
to ICTP and its associate office for their hospitality where the
main parts of this work were conducted and finally finalized.

%@@@@@@@@@@@@@@@@@@@@@@@@@@@@@@@@@@@@@@@@@@@@@@@@@@@@@@@@

%@@@@@@@@@@@@@@@@@@@@@@@@@@@@@@@@@@@@@@@@@@@@@@@@@@@@@@@@

\end{document}